\documentclass[compsoc,conference,a4paper,10pt]{IEEEtran}
\IEEEoverridecommandlockouts

\usepackage{cite}
\usepackage{amsmath,amssymb,amsfonts}
\usepackage{algorithmic}
\usepackage{graphicx}
\usepackage{textcomp}
\usepackage{xcolor}
\usepackage{tikz}
 \usepackage{booktabs}
\usetikzlibrary{arrows.meta,positioning,shapes.geometric}
\usepackage{adjustbox} 
\tikzset{>=Latex}
\usepackage{lipsum}
\usepackage[colorlinks=true,urlcolor=black]{hyperref}
 \usepackage[caption=false,font=footnotesize]{subfig}
 \usepackage{array} 
\usepackage[utf8]{inputenc}
\def\BibTeX{{\rm B\kern-.05em{\sc i\kern-.025em b}\kern-.08em
    T\kern-.1667em\lower.7ex\hbox{E}\kern-.125emX}}
\usepackage{enumitem}

\begin{document}

\title{Fuzz'EMup: Leveraging EM Side-Channel Emanation to Guide Black-Box Embedded Firmware Fuzzing}

\author{
\IEEEauthorblockN{Fatemeh Moradihaghighi\textsuperscript{*}}
\IEEEauthorblockA{\textit{School of Computing}\\
Clemson University\\
fmoradi@clemson.edu}
\and
\IEEEauthorblockN{Zihao Zhan\textsuperscript{*}}
\IEEEauthorblockA{\textit{Department of Computer Science}\\
Texas Tech University\\
zihao.zhan@ttu.edu}
\and
\IEEEauthorblockN{Yanan Guo}
\IEEEauthorblockA{\textit{Department of Computer Science}\\
University of Rochester\\
yguo51@cs.rochester.edu}
\and
\IEEEauthorblockN{Ziming Zhao}
\IEEEauthorblockA{\textit{Khoury College of Computer Sciences}\\
Northeastern University\\
z.zhao@northeastern.edu}
\and
\IEEEauthorblockN{Mashrur Chowdhury}
\IEEEauthorblockA{\textit{Department of Civil Engineering}\\
Clemson University\\
mac@clemson.edu}
\and
\IEEEauthorblockN{Zhenkai Zhang}
\IEEEauthorblockA{\textit{School of Computing}\\
Clemson University\\
zhenkai@clemson.edu}
}
\maketitle

\begingroup
\renewcommand\thefootnote{*}
\footnotetext{These authors contributed equally to this work.}
\endgroup\

\begin{abstract}
As IoT and embedded devices proliferate across various domains, securing their firmware has become critical. 
Fuzzing offers a systematic approach to uncovering vulnerabilities in firmware, and coverage feedback can improve its effectiveness by guiding exploration. 
However, many devices make coverage information impossible to obtain by preventing firmware extraction, instrumentation, or accurate emulation; in such cases, testers are left with only inefficient black-box fuzzing. 

In this paper, we present an approach that leverages electromagnetic (EM) side-channel emanations to guide firmware fuzzing in purely black-box settings. 
However, turning raw EM measurements into reliable guidance is challenging: EM traces are noisy, and timing jitter causes corresponding features in different traces to shift in time. 
We address these challenges by combining frequency band selection based on the activity-to-idle signal contrast with dynamic time warping to align per-input traces and detect sustained divergence, while maintaining scalability by organizing executions in a tree structure based on their divergence times. 
We evaluate our approach on four real firmware targets and demonstrate that EM-derived feedback enhances path exploration, yielding higher code coverage than unguided fuzzing. 

\end{abstract}

\section{Introduction}
\label{sec:introduction}
IoT and embedded devices are widely deployed across consumer and industrial environments. 
Their firmware is a key component that has full control over these systems. 
Consequently, bugs in firmware may be exploited to cause serious security breaches and, given that many embedded devices are safety-critical cyber-physical systems, result in physical harm or loss of life. 
Despite firmware developers' efforts to write secure code, bugs continue to be discovered in deployed firmware~\cite{muench2018you}. 
This reality demands systematic firmware testing. 
Among the available approaches to testing firmware, fuzzing has proven particularly effective in uncovering potential bugs~\cite{zhu2022roadmapSurvey}. 

In essence, fuzzing is an automated technique for exploring program behavior by executing the target with numerous generated inputs, including malformed or unexpected ones, while monitoring for crashes, hangs, or other anomalous behavior~\cite{sutton2007fuzzing}. 
For efficiency, fuzzing is often based on mutation, which maintains a corpus of seed inputs and derives new test cases by mutating these seeds~\cite{AFL}. 
To guide the mutation process toward more productive exploration, execution feedback is often used to determine which seeds to retain in the corpus and prioritize for further mutation, a strategy known as guided fuzzing. 
In practice, coverage-guided fuzzing is the most widely adopted form of guided fuzzing, which leverages code coverage as the feedback signal to identify promising inputs. 
Coverage feedback is typically obtained by dynamic binary instrumentation~\cite{mera2024shift, clements2020halucinator, gustafson2019toward} or by rehosting the binary in emulation frameworks that expose coverage counters~\cite{maier2019unicorefuzz, newsome2005dynamic}. 
Both instrumentation-based and emulation-based coverage collection assume that the firmware binary can be extracted and either modified or executed in a controllable environment. 
However, such assumptions frequently fail due to several factors. 

First, many devices actively prevent firmware extraction through hardware protection mechanisms. 
For instance, STM32 microcontrollers provide read-out protection, Microchip AVR devices offer code protection fuses \cite{ling2017iotsecurity}, and TI MSP430 controllers feature debug lockdown capabilities, all of which block access to the firmware image through debug interfaces. 
Second, even when firmware extraction succeeds, the binary may be encrypted or signed, rendering it unmodifiable and its internal structure opaque. 
Examples include flash encryption in ESP32 devices and secure boot in NXP i.MX processors, which prevent both modification and analysis of the firmware code.
Third, some devices employ proprietary processor architectures, such as Apple's secure enclave processor or NVIDIA's Falcon \cite{zhu2017discreteGPUs}, for which no emulation environment exists, making rehosting infeasible regardless of whether the firmware can be extracted.

In these scenarios, testers can only treat the target device as a black box and fuzz its firmware through random interactions with it. 
Undoubtedly, such black-box fuzzing is significantly less effective at systematically exploring execution paths and can readily miss bugs that coverage-guided methods would expose. 
Accordingly, we ask the following research question: 
\emph{Can we derive a coverage proxy to guide firmware fuzzing in purely black-box settings where only external device interactions are observable?}

In this work, we answer this question affirmatively by leveraging physical side-channel signals, which are inevitably emitted during device operation, to infer firmware control-flow behavior for guiding fuzzing. 
Specifically, we use electromagnetic (EM) emanations as our feedback signal. 
EM traces offer several advantages: (1) they can be captured non-invasively with external probes, requiring no device modification; (2) they exhibit high sensitivity to instruction execution and control-flow changes; and (3) they provide sufficient temporal resolution to detect fine-grained execution differences. 
Nevertheless, leveraging EM side-channel information for firmware fuzzing presents several challenges. 

First, raw EM traces are noisy. 
Control-flow-correlated signals are typically mixed with unrelated emissions, making it difficult to reliably infer execution behavior. 
Additionally, timing variations across runs (e.g., due to microarchitectural effects and asynchronous activity) cause corresponding features in different traces to shift in time, even when the underlying execution path is identical. 
To address these issues, we filter each captured EM trace to an informative frequency band and convert it into an execution-sensitive signal. 
We then align the filtered trace with those from previous executions using Dynamic Time Warping (DTW), which finds a non-linear time mapping between signals, allowing similar patterns to align despite local timing variations. 

Second, we must determine when an input explores a genuinely new execution path rather than merely producing a minor perturbation of an existing one. 
Fortunately, DTW provides additional utility beyond alignment. It yields a warping path that reveals where the alignment stretches or compresses time. 
We analyze this warping path to compute an approximate divergence point along the aligned traces. 
When two executions follow the same control-flow path, their index offset along the warping path remains nearly constant with only small fluctuations due to timing jitter. 
In contrast, when control flow diverges, the offset exhibits a sustained drift. 
We use this divergence point as a coverage-like signal, and an input is considered to explore a new path only when its trace exhibits sustained divergence from all previously observed executions. 

Third, comparing traces at scale poses a computational challenge. 
Directly comparing each new trace against all prior traces is expensive and quickly becomes a throughput bottleneck as the corpus grows. 
To achieve scalability, we organize execution traces in a tree structure that groups them by divergence time. 
Each new trace is compared only to a small number of representative traces along a single root-to-leaf path rather than against the entire corpus. 
This tree structure both filters redundant executions and enables efficient seed selection strategies that prioritize inputs likely to discover new coverage.

With these challenges addressed, we implement an EM-guided embedded firmware fuzzer and evaluate it on four real firmware targets. 
Our results demonstrate that EM-guided fuzzing consistently outperforms unguided fuzzing. 
It achieves higher code coverage with fewer inputs, discovers paths that the unguided approach misses, selects coverage-increasing seeds 1.6$\times$ more frequently, and maintains scalable performance as the corpus grows. 

In summary, this work makes the following contributions:
\begin{itemize}[nosep, leftmargin=*]
\item 
We present the first EM-guided approach for black-box embedded firmware fuzzing, demonstrating that such side-channel information can serve as a coverage proxy and enable guidance without requiring binary extraction, instrumentation, or accurate emulation. 

\item 
We develop a signal processing pipeline that filters, aligns, and compares EM traces to detect sustained divergence, enabling reliable identification of inputs that trigger novel execution behaviors despite noise and timing jitter. 

\item 
We introduce a divergence tree structure that organizes executions by their divergence times, enabling scalable EM trace comparison and seed selection strategies. 

\item 
We evaluate our approach on four firmware targets and show that EM guidance achieves consistent improvements in coverage, input efficiency, and path discovery compared to unguided fuzzing. 
\end{itemize}

\section{Background}

In this section, we review how state-of-the-art coverage-guided fuzzers utilize feedback to guide input selection. 
We then introduce EM side-channel analysis as a commonly used method to observe computations on embedded devices.
\subsection{Coverage-guided Fuzzing}

Fuzzing is an automated testing technique that repeatedly executes a program with generated inputs to trigger failures and expose bugs. 
Modern fuzzers are often categorized by the amount of internal program information they use. Black-box fuzzers rely only on externally visible behavior, white-box fuzzers use heavyweight program analysis (e.g., symbolic execution), and greybox fuzzers use lightweight feedback signals collected during execution \cite{rawat2017vuzzer}. Coverage-guided fuzzing is the dominant greybox approach in practice. It uses code-coverage feedback to decide which inputs to keep and mutate further, with the goal of systematically expanding explored behaviors and reaching deeper program states \cite{bohme2017directed}.

A coverage-guided fuzzer maintains a corpus of interesting inputs. 
Iteratively, it selects a seed from the corpus, applies mutations, executes the target on each mutated input, and collects feedback. 
If an input increases the chosen coverage metric, it is promoted into the corpus and becomes a new mutation starting point; otherwise it is discarded. 
In parallel, the fuzzer records crashes and other abnormal terminations for later triage  \cite{li2018fuzzing}. This feedback loop converts largely random mutation into a search process biased toward inputs that measurably expand explored execution behavior.

In most coverage-guided fuzzers, the feedback signal is a compact representation of the control flow exercised by an input. A common implementation is a lightweight coverage map recorded in a shared bitmap, designed to keep per-execution overhead low while still providing a practical novelty signal \cite{AFL, fioraldi2020afl++}. Coverage can be defined at different granularities, such as function, basic-block, or edge, and fuzzers differ in what they treat as new coverage and how they aggregate it \cite{bohme2016coverage}. In all cases, coverage serves as an efficiently computable proxy for behavioral novelty that determines which inputs are retained and prioritized.

Obtaining conventional coverage feedback typically requires instrumentation or tracing. 
When source code is available, fuzzers often rely on compile-time instrumentation to update coverage metadata efficiently \cite{fan2020arm, song2019periscope}. 
When only a binary image is available, coverage can sometimes be recovered using dynamic binary translation or hardware tracing~\cite{delshadtehrani2020phmon}, but these mechanisms are not uniformly available and can incur substantial overhead. 
For embedded firmware, coverage is commonly obtained either by instrumenting the firmware image directly~\cite{oh2015less} or by rehosting it in an emulator and instrumenting the emulated CPU~\cite{feng2020p2im, mera2021dice, chen2016Firmadyne, zheng2019firm}. In black-box embedded firmware fuzzing on real microcontroller units (MCUs), such instrumentation-dependent options are often unavailable, so conventional coverage feedback cannot be assumed~\cite{chen2018iotfuzzer}. An alternative feedback signal is therefore required to drive corpus growth in the same fuzzing loop.

\subsection{EM Side-channel Analysis}

Electromagnetic (EM) side channels arise because digital computation involves time-varying currents in transistors and interconnects, which produce EM emissions from the chip. Such emissions can carry information correlated with firmware program activity and other internal operations, even without software-level observability of the running system \cite{zajic2014experimentalEM, zhang2020rowhammer, zhan2020bitjabber, zhan2022graphics}. 
Because EM can be captured externally with a probe, it provides a non-invasive measurement channel that does not require firmware knowledge or debug privileges, and thus remains viable in constrained-access settings \cite{han2017watch,callan2016zero}.

In practice, EM side-channel analysis treats each execution as a time-series waveform and analyzes how traces change across executions. 
A typical workflow is to repeatedly run the target under controlled inputs, record traces, and compare traces using similarity or distinguishability measures. 
While classical side-channel attacks use such traces to infer sensitive values~\cite{strobel2015scandalee, genkin2016ecdh, genkin2016ecdsa}, the same measurement pipeline can also distinguish internal behaviors across executions based on trace-level differences~\cite{callan2014Savat, callan2015fase, nazari2017eddie}. 

A central practical challenge for EM side-channel analysis is signal variability. EM traces are noisy and can vary across runs due to environmental interference, device operating conditions, and measurement setup. Timing variability is particularly important. Repeated executions can appear shifted or stretched because of interrupts, wait states, peripheral interactions, or input-dependent delays. This makes naive point-wise comparisons unreliable and motivates alignment or comparison methods robust to temporal misalignment. 
As a result, practical EM side-channel analysis typically relies on signal processing and comparison techniques designed to tolerate noise and timing variation.~\cite{sehatbakhsh2019remote}

\section{Overview}
In this section, we first describe the threat model. We then present a high-level overview of the components of our design.
\subsection{Threat Model}
This work targets settings where the firmware binary and debug access is unavailable. 
In such black-box conditions, conventional coverage-guided fuzzers cannot obtain code-coverage feedback because binary instrumentation is impossible and accurate rehosting is infeasible. 
We make no assumptions about knowledge of the source code, symbols, debug privileges, board schematics, or vendor documentation. 
Inputs are delivered only through deployed peripheral interfaces, and observability is limited to externally visible outputs and non-invasive side-channel measurements. 
We assume a tester can execute the firmware repeatedly with chosen inputs and position the probe at arbitrary locations near the MCU package, but cannot modify the firmware or obtain internal execution visibility. 

\subsection{Design}

\begin{figure}[!ht]
\centering
  \includegraphics[width=\linewidth]{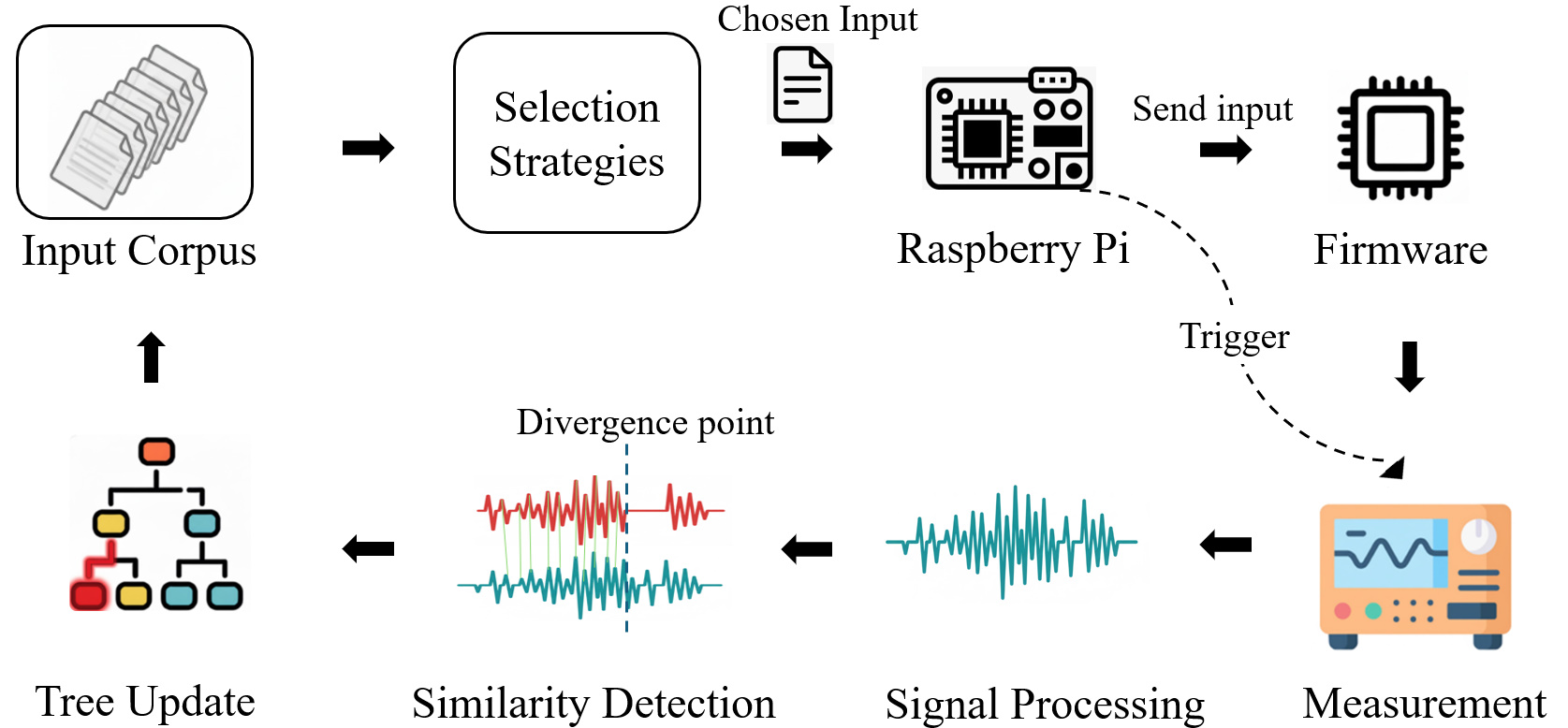}%
 \caption{Engine sends inputs to the Pi; the Pi forwards via UART to the MCU and issues a trigger to the scope (dashed). The MCU’s EM emission is captured by the scope; traces are consumed by the engine.}
\label{fig:probe-fuzz-arch}
\end{figure}

The overall framework is shown in Fig.~\ref{fig:probe-fuzz-arch}. The host selects a seed from the corpus using a selection strategy, mutates it, and sends the input to a Raspberry~Pi bridge, which forwards it via the device's input interface while simultaneously triggering EM capture. In principle, the bridge could also record conventional black-box signals (e.g., resets, error codes, or timeouts) returned by the MCU, but in this work we do not use such signals as feedback and base all guidance exclusively on the EM trace.
An H-field probe over the MCU records the waveform. The signal is preprocessed to isolate the active execution window. 

Our method maintains a tree-structured clustering of prior traces, where each node stores a representative trace and corresponds to a family of similar executions. The new trace is compared iteratively as the structure is traversed. At each visited node, it is DTW-aligned with the node’s representative trace, and the alignment is used to compute an approximate divergence time, i.e., the earliest point at which the trace departs from the representative trace. This estimate determines which branch to follow next. If the trace exhibits a sustained divergence that does not match any existing branch, a new node is created and the corresponding input is added to the corpus; otherwise, it is treated as similar behavior and the fuzzing loop continues without modifying the tree. The resulting divergence tree both filters redundant executions and serves as the state used by the seed selection policies in Section~\ref{sec:selection}

\section{Implementation}
\label{sec:implementation}
The core assumption behind our approach is that the EM traces correlate with the device's computation. When the firmware follows different control-flow paths, the on-chip current draw and switching activity change, producing distinguishable patterns in the measured EM emissions. By comparing EM traces from different executions, we obtain a control-flow–sensitive signal that can serve as a coverage proxy for guiding fuzzing.
However, using EM side-channel traces to guide fuzzing introduces several practical challenges. 

\noindent\textbf{Challenge 1: noisy and unstable EM signal.}
Even when the device executes the same input repeatedly, the measured traces are not perfectly identical.
The variation comes from two sources.
First, amplitude noise arises because the measuring probe inevitably picks up signals from both the target components and other irrelevant components on the board, as well as environmental RF interference. The overlap of these sources with the signal of interest lowers the effective signal-to-noise ratio (SNR).
Second, timing variability arise as the timing of operations can vary between runs due to microarchitectural effects and asynchronous activity, so corresponding features in different traces may be shifted in time. This jitter and misalignment make direct trace comparison difficult, even when the underlying execution path is the same.

\noindent\textbf{Challenge 2: detecting path divregence.}
To use EM traces as a coverage proxy, we need to determine where two executions diverge in their control-flow paths, so that we can tell whether a new input truly explores a new path. This is non-trivial because executions that follow the same control flow can still produce slightly different EM traces, so naive distance thresholds or pointwise comparisons risk mistaking noise, interrupts, or microarchitectural jitter for genuinely new behavior. We therefore need a divergence indicator that is robust to small fluctuations yet sensitive to sustained, path-level differences along the trace. 

\noindent\textbf{Challenge 3: comparing traces at scale is computationally expensive.}
Fuzzing can generate a large corpus of executions, and using a robust alignment-based similarity measure between every new trace and all existing traces quickly becomes a bottleneck. Algorithms that account for local time distortions are typically more costly than simple pointwise distances. For example, time warping with a Sakoe-Chiba band of half-width $W$ over traces of length $M$ costs $O(MW)$ per comparison, so the per-trace comparison cost grows rapidly with both trace length and corpus size. This overhead directly impacts fuzzing throughput and limits the extent to which EM-based feedback can be used in the input selection loop.

This section details how we address these challenges by combining band selection and execution-window detection with DTW-based alignment to compare EM traces robustly despite jitter and broadband noise, and to use divergence as an indicator that an input has triggered a new execution path.

\subsection{Signal Acquisition and Activity Selection}

Our method exercises the firmware via an existing input interface so that fuzz inputs are delivered as normal device traffic. A Raspberry Pi bridge coordinates input injection and oscilloscope triggering to synchronize EM trace capture with each test input.
To maximize the SNR, we measure EM signal at a probe position and frequency band chosen to jointly maximize it. 

\noindent\textbf{Localized EM measurement.} To find the location with the highest SNR, a near-field probe is positioned close to the microcontroller and scanned across the board to sample candidate locations. At each location, we record short traces in both idle and active states and measure the change in EM activity between them. We then select a probe position where execution causes a strong, repeatable increase over the idle baseline while minimizing coupling to unrelated components.

\noindent\textbf{Frequency band selection.}
To select a frequency band, we compute a spectrogram using the STFT and examine how strongly each frequency component separates idle from active behavior. Let $S(f,t)$ denote the STFT and $E_f(t) = |S(f,t)|$ its magnitude at frequency $f$. For each frequency $f$, we compute the Welch $t$–statistic using Eq.~\ref{eq:welch}, which measures the standardized difference in mean $E_f(t)$ between active and idle regions, reflecting how informative this frequency band is.
\begin{equation}\label{eq:welch}
t(f) \;=\;
\frac{\mu_{\text{act}}(f) - \mu_{\text{idle}}(f)}
     {\sqrt{\dfrac{\sigma_{\text{act}}^2(f)}{n_{\text{act}}}
            + \dfrac{\sigma_{\text{idle}}^2(f)}{n_{\text{idle}}}}}\,,
\end{equation}

$\mu_{\text{act}}(f)$ and $\mu_{\text{idle}}(f)$ are the sample means of $E_f(t)$ when the device is active and idle at frequency $f$, $\sigma_{\text{act}}^2(f)$ and $\sigma_{\text{idle}}^2(f)$ are the corresponding sample variances, and $n_{\text{act}}$, $n_{\text{idle}}$ are the number of samples in each case. Frequencies with larger $|t(f)|$ provide a clearer separation between active and idle behavior and thus higher effective SNR for execution-related activity.

\begin{figure}[ht]
  \centering
  \includegraphics[width=\linewidth]{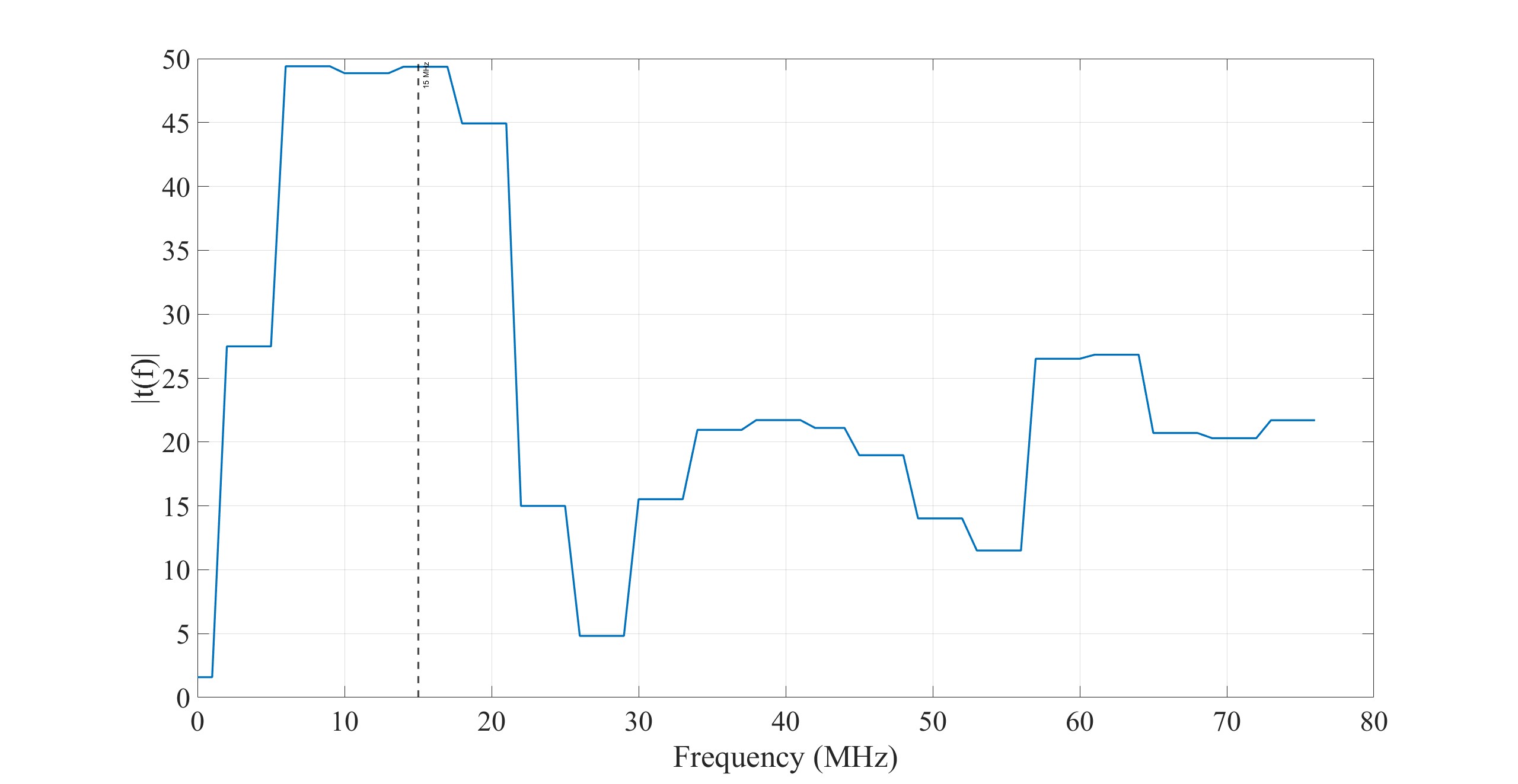}%
  \caption{Welch statistic \(|t(f)|\) over frequency for a representative trace.
  The peak near 15\, MHz marks a band with strong separation between idle and active states.}
  \label{fig:metric}
\end{figure}

Figure~\ref{fig:metric} plots $|t(f)|$ for a representative trace. The
statistic attains its maximum near 15\,MHz, indicating that this
frequency best distinguishes active and idle behavior; we therefore
select 15\, MHz as the carrier for subsequent demodulation. This procedure is run once per device/firmware target on a small calibration set, and the resulting carrier band is then fixed for all subsequent fuzzing runs on that device.

\begin{figure}[ht]
  \centering
  \includegraphics[width=\linewidth]{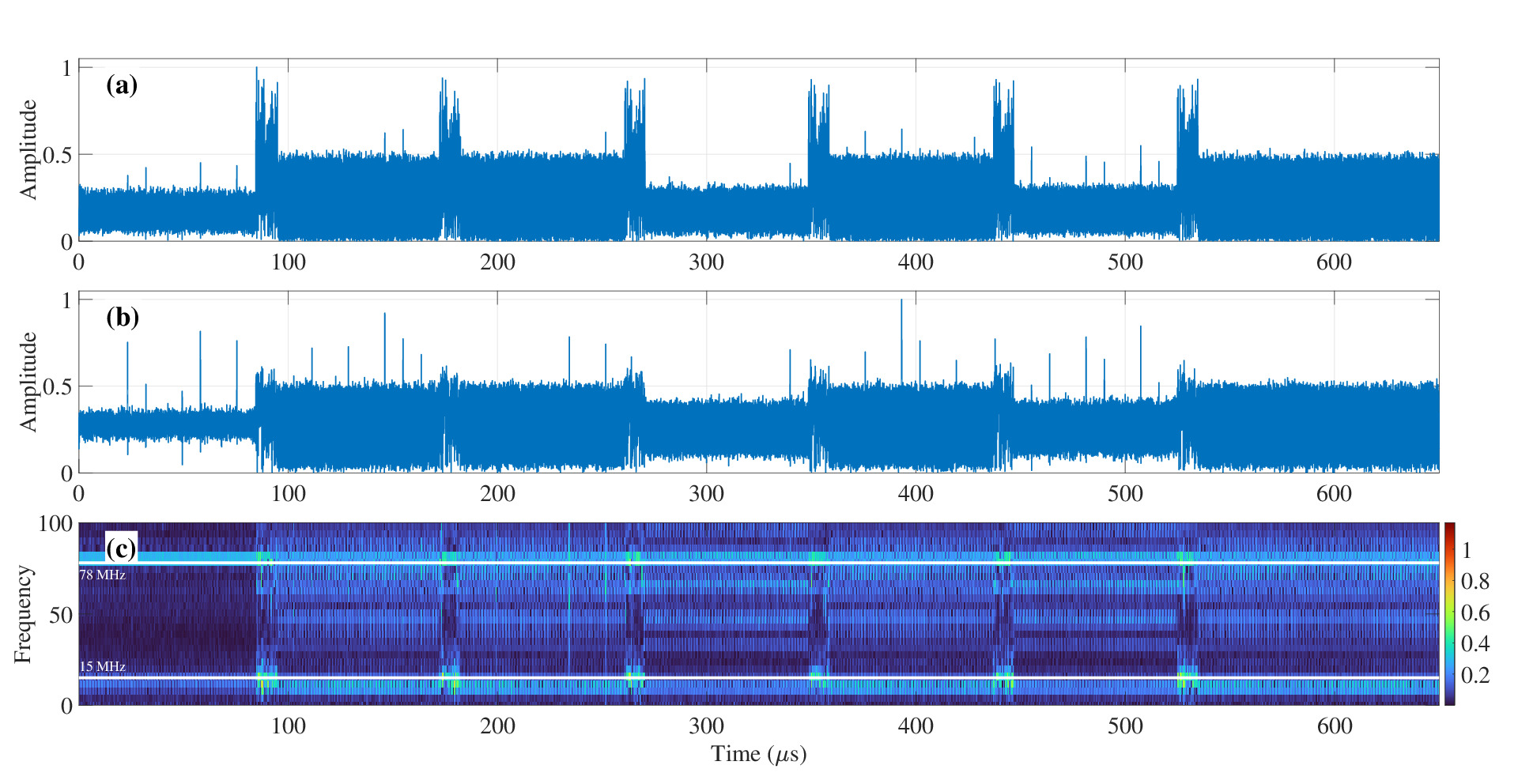}%
  \caption{Demodulated envelopes at (a) 15\,MHz and (b) 78\,MHz, and (c) the corresponding STFT spectrogram. The 15\,MHz band shows a clearer rise during execution and higher contrast over idle.}
  \label{fig:stft}
\end{figure}

To visualize the effect of this selection, Figure~\ref{fig:stft} shows the demodulated time-domain envelopes at two carrier frequencies along with the corresponding spectrogram. These bands show different time-domain behaviors. Both carriers are drawn from regions of the spectrum with high energy, but they differ in how well they encode execution activity. Among such high-energy bands, the selection criterion is to prefer carriers whose demodulated envelopes exhibit strong contrast between active and idle intervals and a stable, repeatable waveform shape across executions. The 15\, MHz component exhibits a clear rise during execution and a strong contrast relative to idle, whereas the 78\, MHz component is dominated by background fluctuations and the execution-related bursts are less pronounced, so the correlation with the activation window is weaker and the underlying shape is less well preserved.

\noindent\textbf{Execution–window detection.}
Given the selected frequency band $f_c$, we project each trace $x(t)$ onto the complex sinusoid at that carrier and take the magnitude to obtain a band-limited activity envelope. Concretely, we define
\begin{equation}
  h(t) = e^{j 2\pi f_c t},
  \qquad
  e(t) = \bigl|(x * h)(t)\bigr|,
\end{equation}
where $*$ denotes convolution.  This operation acts as a short complex demodulation that emphasizes energy near $f_c$. After temporal smoothing and down-sampling, $e(t)$ yields a smooth activity signal that preserves the coarse shape of the execution while suppressing high-frequency noise. An idle baseline level is estimated from low-energy portions of $e(t)$, and the execution interval is defined as a sustained region where $e(t)$ remains clearly above this baseline. This produces a robust, low-parameter estimate of the active window that does not rely on precise trigger timing.

\subsection{DTW-based Trace Comparison}
Dynamic Time Warping (DTW) is a standard technique for aligning non-linear time series. It finds an optimal correspondence between two sequences that may vary in local timing, and is widely used for speech recognition and general time-series matching~\cite{sakoe2003dynamic}. DTW has also been applied in side-channel analysis to elastically realign noisy traces before statistical comparison~\cite{van2011improving}.

Let $r = (r_1,\dots,r_n)$ denote the reference trace and $c = (c_1,\dots,c_m)$ the candidate trace. DTW constructs an $n \times m$ cost matrix whose entry $(i,j)$ stores the local distance $d(r_i,c_j) = |r_i - c_j|$. An alignment between $r$ and $c$ is represented by a \emph{warping path} $P = (p_1,\dots,p_K)$, where each $p_k = (i_k,j_k)$ indexes a matrix cell and $\max(n,m) \le K \le n + m - 1$. A valid warping path starts in the top-left corner $(1,1)$ and ends in the bottom-right corner $(n,m)$, moves in unit steps (to $(i+1,j)$, $(i,j+1)$, or $(i+1,j+1)$), and never moves backwards along either axis, so the indices are non-decreasing and time is preserved in both sequences. Intuitively, the path specifies how samples in the candidate trace are stretched or compressed to align with samples in the reference trace.

Among all valid paths, DTW chooses the one with the minimum accumulated cost,
\begin{equation}
{DTW}(r,c) = \min_{P} \sum_{(i,j)\in P} d(r_i, c_j),  
\label{eq:dtw}
\end{equation}
and the optimal path is obtained via standard dynamic programming over the cost matrix. 
Searching the full $n \times m$ matrix is often unnecessary and can produce unrealistic alignments, so we constrain the search using a Sakoe--Chiba band, which restricts the path to lie within a diagonal corridor around the line of uniform time alignment. Concretely, for a point $(i,j)$ on the warping path we require
\[
  \frac{n}{m}\,i - W \;\le\; j \;\le\; \frac{n}{m}\,i + W,
\]
where $W$ is the half-width of the band. This bounds how far the candidate index $j$ may drift from the linearly scaled reference index $(n/m)i$, limiting local time distortion while reducing computational cost from $O(nm)$ to $O(W \cdot \max(n,m))$. In our setting, it is chosen to be large enough to cover typical jitter with a safety margin, but small enough to keep the band within a narrow corridor around the diagonal. We use banded DTW to realign EM traces under modest timing jitter before estimating divergence points.

\begin{figure}[ht]
  \centering
  \includegraphics[width=\linewidth]{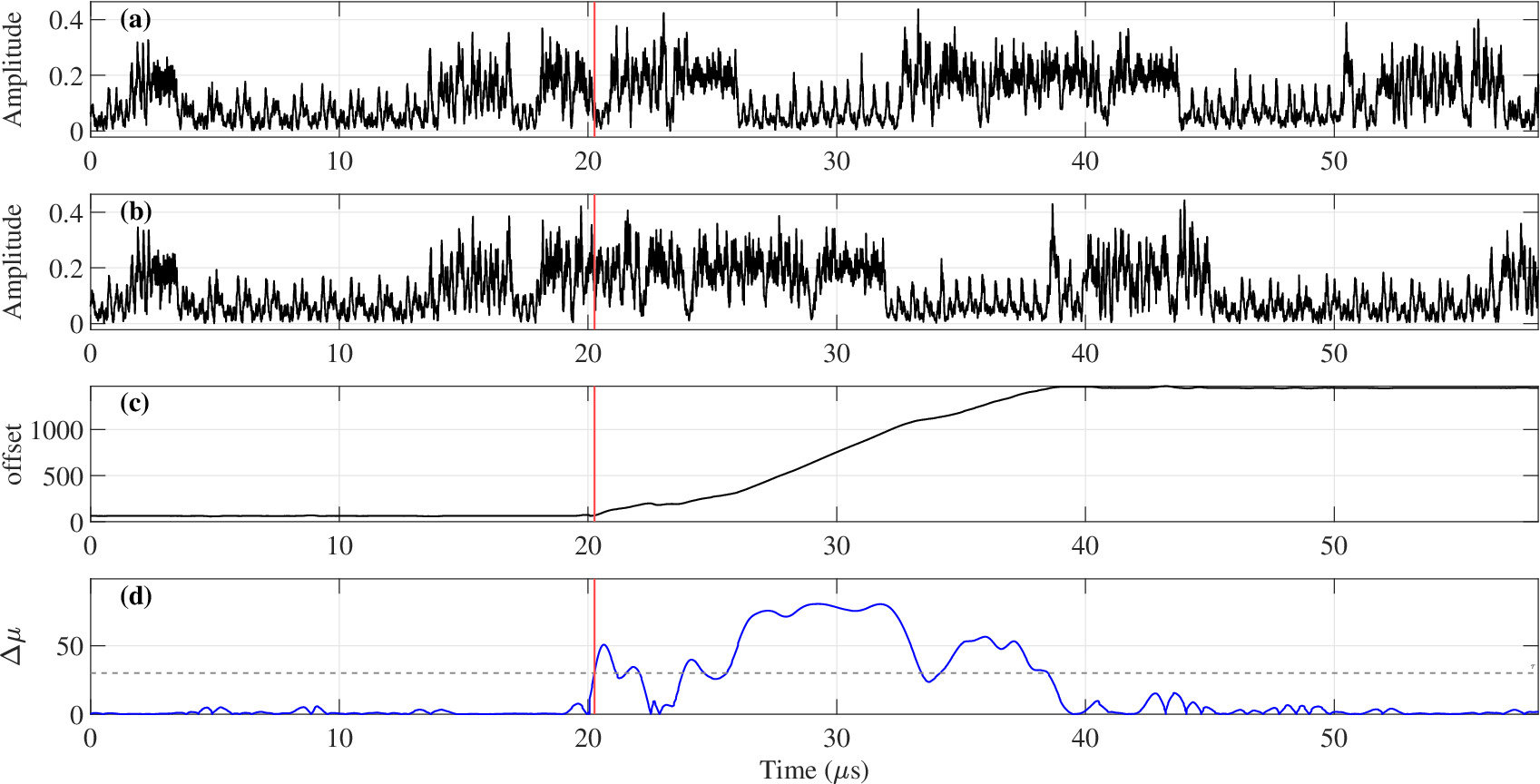} 
  \caption{DTW-aligned divergence example.
  (a–b) Aligned reference and candidate traces with estimated divergence
  times (red). (c) DTW offset profile $\mathrm{offset}[k]$. 
  (d) Local detector $\Delta\mu[k]$ used to pick the divergence index.}
  \label{fig:dtw-div}
\end{figure}

From the warping path, we derive an offset profile $\mathrm{offset}[k] \;=\; i_k - j_k$ over the path
positions $k = 1,\dots,K$ where $i_k$ and $j_k$ are the reference and candidate indices aligned at step $k$ on the DTW path. This offset captures the local index shift between the
two traces. When two executions follow the same control-flow path, $\mathrm{offset}[k]$ remains nearly constant apart from small timing jitter; when their control flow diverges, the alignment must stretch one trace relative to the other, and $\mathrm{offset}[k]$ exhibits a sustained drift rather than short-lived fluctuations. We detect such drift using a simple local change detector. For each position $k$ along the path where both windows fit (e.g., $L \le k \le K-L$), we compare the mean offset in a pre-window and a post-window of length $L$:
\[
  \Delta\mu[k]
  \;=\;
  \frac{1}{L}\sum_{u=k+1}^{k+L} \mathrm{offset}[u]
  \;-\;
  \frac{1}{L}\sum_{u=k-L}^{k-1} \mathrm{offset}[u].
\]
Here, $L$ is a small window length that controls the smoothing scale of the detector. We then mark the earliest index $k_{\mathrm{div}}$ at which $\Delta\mu[k]$ exceeds a threshold $\tau$ for a sustained region and take the corresponding times in the reference and candidate traces as the estimated divergence point. 
Figure~\ref{fig:dtw-div} illustrates this procedure. Panels~(a) and~(b) show the reference trace $r$ and candidate trace $c$ over the aligned time window; the vertical lines indicate the estimated divergence times $t_{\mathrm{div},r}$ and $t_{\mathrm{div},c}$ in each trace. Panel~(c) plots the smoothed offset profile
$\mathrm{offset}[k]$ along the DTW warping path over the same region; the change in slope around $k_{\mathrm{div}}$ indicates that the reference and candidate indices no longer advance at the same rate, consistent with a divergence in control flow. 
Panel~(d) shows the detector $\Delta\mu[k]$ and the threshold $\tau$: once $\Delta\mu[k]$ crosses $\tau$ and remains above it, we mark the earliest such index as $k_{\mathrm{div}}$, which corresponds to the divergence times highlighted in panels~(a) and~(b).

\subsection{Divergence Tree Trace Indexing}
A naive uniqueness test needs to compare each new trace against \emph{all} previous traces, yielding $O(N)$ alignments per insertion (and $O(N^2)$ total), with each alignment itself costing $O(MW)$ for length $M$ and band $W$. 
To avoid this quadratic growth, we cluster traces by their earliest sustained divergence time and organize them in a divergence tree. 
Executions that share control flow up to time $t^\star$ tend to align well with one another and poorly with traces whose first branch occurs far earlier or later. 
When inserting a new trace, we only align against a small set of \emph{representatives} along a single root-to-leaf path (whose depth is $\ll N$), so the time per insertion is approximately $O(\mathrm{depth}\cdot MW)$ instead of $O(NMW)$. 

In practice, the depth grows sublinearly, so search and insertion are substantially cheaper than flat all-pairs comparison, while preserving the ability to discover novel behaviors at their branch points. Because each new sustained divergence either descends into an existing child with a nearby divergence time or creates a new sibling, any behavior whose earliest divergence time lies outside all existing child windows will be routed to a new node, so genuinely new paths tend to create new branches rather than being merged into existing ones. 
This divergence tree serves as a practical index for selection, with each node recording its DTW divergence time. New traces are routed by approximate divergence time rather than compared against all prior traces. Traces that agree up to time \(t^\star\) follow the same path to that depth, and a newly detected sustained divergence creates a child at the corresponding time.

\begin{itemize}
  \item \textbf{Node contents.} Each node stores its divergence time relative to the parent and a representative trace for the branch. The representative is simply the first trace whose
execution created that node.
  \item \textbf{Insertion.} For a new trace, we start at the root and compare it against the current node’s representative. If no sustained divergence is detected at some node, we treat the trace as redundant and stop; otherwise, we record the divergence time $t^\star$ and either descend or create a child as described below. 
  \item \textbf{Descent by proximity.} If $t^\star$ is near an existing child's divergence time, we descend into that child and repeat the comparison; otherwise, we create a new sibling at the current level with divergence time $t^\star$ and add the trace to the seed
pool.
\end{itemize}

This process yields a tree whose structure reflects where inputs tend to branch, with proximity in divergence time guiding insertion. 
For each device, we instantiate this proximity as a small microsecond-scale window around each node’s divergence time, derived from its core clock frequency and EM trace time resolution (sampling rate, downsampling, smoothing), so that nearby divergences correspond to only a few hundred instructions of execution.

\begin{figure}[ht]
    \centering
    \includegraphics[width=\linewidth]{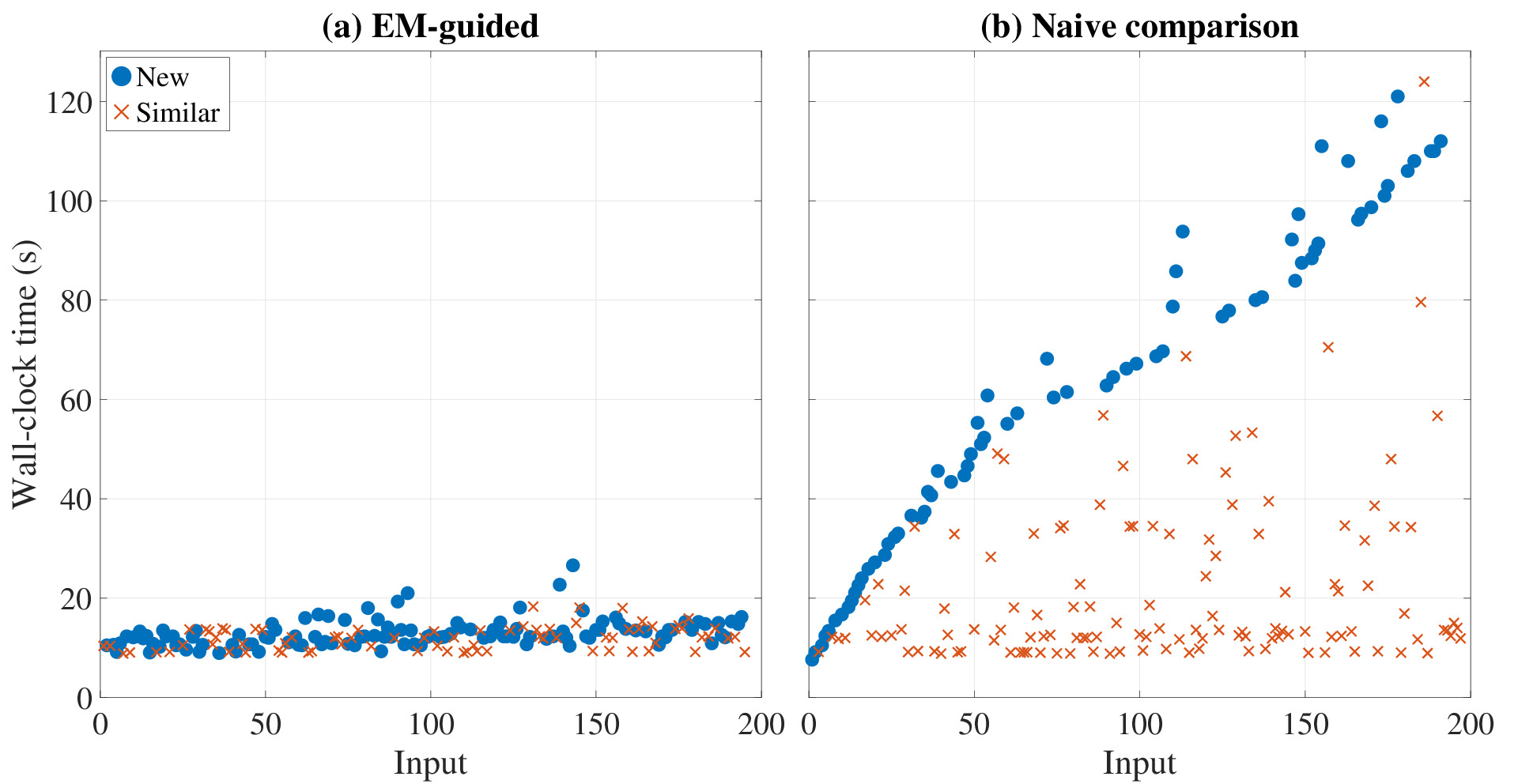}
    \caption{Per-input wall-clock time for (a) EM-guided clustering and (b) a naive flat baseline on one firmware target.}
    \label{fig:tree-runtime}
\end{figure}

To demonstrate the scalability benefit of divergence tree trace indexing, we compare the per-input trace classification time of our method against a flat baseline that aligns each new trace to all existing corpus representatives. 
Fig.~\ref{fig:tree-runtime} plots the wall-clock time required to complete these comparisons and classify each candidate trace as either similar to an existing representative or representing a new execution path. In the flat baseline, per-input time increases approximately linearly as the corpus grows, for both candidates ultimately deemed similar and candidates that become new representatives, because each insertion performs $O(N)$ alignments. In contrast, the divergence-tree method remains approximately constant because it performs comparisons only along a single root-to-leaf path whose depth grows much more slowly than $N$.

Having distinguished similar versus novel executions and organized them into a searchable index, we next address \emph{which node to mutate}. Section~\ref{sec:selection} details four complementary selection strategies—most descendants, most immediate children (high out-degree), frontier leaves, and random—together with an $\varepsilon$-greedy + Upper Confidence Bound (UCB) scheduler that mixes them.

\subsection{Node Selection and Scheduling}\label{sec:selection}

Prior work~\cite{gan2018collafl} shows that seed selection strongly shapes a fuzzer’s path exploration and input quality, yet choosing optimal seeds is hard and can leave parts of the execution space unvisited. 
We address this “what to mutate” problem with four complementary node-selection strategies. Strategies 1–4 respectively target branch hubs, local split points, the frontier, and stochastic diversity. Each strategy selects one node from a top-$k$ shortlist to mutate, thereby avoiding over-concentration on a single hub.

\begin{enumerate}
  \item \textbf{Most descendants.} Choose the top-$k$ internal nodes with the largest subtree sizes (number of descendants). Such nodes typically lie near major behavioral branch points; mutating them is likely to unlock additional regions.
  \item \textbf{Most immediate children.} Choose the top-$k$ nodes with the greatest number of direct children. High local fan-out indicates a nearby split; mutating at these split points tends to diversify executions.
  \item \textbf{Leaf/frontier.} Among leaves which show nodes without children, choose the top-$k$ by time since last mutation (oldest first). Leaves mark the exploration frontier; prioritizing the oldest avoids repeatedly probing the same endpoints.
  \item \textbf{Random.} Uniformly sample $k$ nodes to inject controlled noise, reduce heuristic bias, and occasionally surface overlooked areas.
\end{enumerate}

In our setting we do not have coverage feedback (edges, basic blocks, or paths), only the execution tree built from EM traces and DTW divergences. The only signal we observe after mutating a seed is whether the resulting input inserts as a new node in the tree (a previously unseen execution path) or is discarded as too similar. We therefore model the strategy selection as a \emph{multi-armed bandit}  problem~\cite{berry1985bandit, yue2020ecofuzz}. Each node-selection strategy is an arm, and each time we use strategy~$i$ we receive a binary reward, $r=1$ if the generated input becomes a new child in the tree and $r=0$ otherwise. The goal of the scheduler is to allocate future mutations to strategies so as to maximize the total number of new nodes discovered with fewer inputs.

To choose among these strategies, we use an $\varepsilon$-greedy + UCB scheduler: with probability $\varepsilon$ we pick a strategy uniformly at random from $\{\mathrm{S1},\mathrm{S2},\mathrm{S3},\mathrm{S4}\}$ (pure exploration); otherwise we select the strategy with the largest UCB score (biased toward exploitation). We maintain, for each strategy $i$:
\begin{itemize}
  \item $N_i$: number of times strategy $i$ has been chosen within the last $W$ choices;
  \item $N=\sum_i N_i$: the total number of strategy choices within the last $W$ choices; 
  \item $R_i=\sum_{j=1}^{N_i} r_{i,j}$: cumulative reward in the last $W$ choices, where $r_{i,j}\in[0,1]$ is the outcome of the $j$-th use of strategy $i$ (we use a binary signal: $r{=}1$ if the generated test inserts as a new child; $r{=}0$ otherwise).
\end{itemize}

The empirical mean reward and the UCB score for strategy $i$ are
\begin{equation}
\label{mu}
\mu_i \;=\; \frac{R_i}{\max(1,\,N_i)} ,
\end{equation}
\begin{equation}
\label{ucb}
\mathrm{UCB}_i \;=\; \mu_i \;+\; c\,\sqrt{\frac{\ln\!\big(\max(1,\,N)\big)}{\max(1,\,N_i)}} ,
\end{equation}
where $c>0$ controls exploration. We use the same definitions later for per-node statistics, with the index $n$ ranging
over tree nodes instead of strategies. Early in a run, when $N_i$ is small, the exploration term in~\eqref{ucb} is large and encourages trying all strategies at least a few times; as more inputs are generated with strategy $i$ and $N_i$ grows, $\mu_i$ stabilizes and the scheduler increasingly exploits the strategy with the largest $\mathrm{UCB}_i$.

After choosing a strategy, that strategy runs its node selector where each strategy builds a top-$k$ candidate set, assigns each node $n$ a primary score according to its heuristic (e.g., most descendants), and sorts candidates in descending order.
For each candidate node $n$, the outcome of its mutations—whether they produce a new child in the tree or not—is treated as a binary reward signal,and a per-node mean reward $\mu_n$ and UCB score $\mathrm{UCB}_n$ are maintained, defined analogously to Eqs.~\eqref{mu} and \eqref{ucb}. We then combine these into a scalar score for each candidate node
\begin{equation}
\mathrm{score}_n \;=\; \mu_n \;+\; \alpha\,\mathrm{UCB}_n  \;-\; \lambda\,\!\left(1 - A_n\right),
\end{equation}
where $A_n$ is the node “age” (generations since the node was last mutated) and $\alpha,\lambda$ are hyperparameters.

We then select the highest-scoring node among the top-$k$ candidates and use its associated input as the seed for the next test.
Because trace acquisition on physical hardware is significantly slower than executing test cases under software emulation, we adapt the mutation stage to this constraint. At each iteration, a mutation stage is randomly chosen and applied to the selected seed. 
We follow AFL-style operators (deterministic/havoc/splice) but use a focused subset: bit flips (1–16 bits), 8/16-bit arithmetic and “interesting” value substitutions, plus a lightweight havoc mode with short stacks of 8/16/32-bit arithmetic, xor/bit-flip edits, small block transforms (rotate/shuffle/duplicate), and optional dictionary inserts. Each new test case applies either one deterministic-style edit or a short havoc-style sequence, rather than a fixed two-phase schedule over the corpus.

Within the chosen seed, the mutation location is primarily selected uniformly at random. To lightly exploit structure in the deterministic stage, we add a small bias toward positions that have historically led to useful behavioral changes. We track, at a coarse granularity, how often mutations at each region of the input resulted in new insertions to the tree, and occasionally choose mutation sites in proportion to this empirical success. This preserves broad exploration while giving slightly more attention to regions that have proven effective in the past.
\section{Evaluations}

We evaluate our method to answer two questions:

\textbf{(Q1) Guidance effectiveness:} Does EM-guided, tree-aware selection reach high coverage faster and uncover behaviors that unguided (random) selection misses under the same mutation operations?

\textbf{(Q2) Novelty efficiency:} Under the same mutation operators and input budget, does EM-guided selection generate a higher fraction of inputs that exercise previously unseen instructions or edges than random selection?

To answer these questions, we compare two fuzzing modes under controlled, matched conditions: (i) \emph{EM-guided, tree-aware selection}, which inserts traces into the divergence tree and schedules seeds using our selection strategies, and (ii) \emph{random unguided selection}, which selects the next seed uniformly at random. Both modes start from the same initial corpus, use the same mutation operators, and are executed concurrently.

Our method itself never directly observes or uses code-coverage feedback. We collect code-coverage measurements only for evaluation to provide a ground-truth measurement of guidance quality. Specifically, for each input executed on hardware we replay the identical input to a rehosted instance of the firmware in QEMU~\cite{bellard2005qemu} and record edge, basic-block, and instruction coverage via QEMU instrumentation. This rehosting infrastructure is not assumed by the threat model and is used solely to compute evaluation metrics. To support the evaluation on our targets, we added a custom STM32L476RG development board model to QEMU, including the Cortex-M4 core, memory map, clock/reset configuration, and the peripherals exercised by our inputs. The rehosted firmware receives the same byte streams as the physical device, delivered to the emulated UART with matching framing, enabling direct coverage measurement for the inputs executed on hardware.

\subsection{Targets and Setup}
We evaluate our approach on four real microcontroller firmware images that run unmodified on an STM32L476RG Cortex-M4 board and exercise different I/O paths and control structures. 
These firmware images have been used in prior work as representative embedded benchmarks~\cite{feng2020p2im, mera2021dice}:
\begin{itemize}
    \item 
    \emph{GPS Receiver.}
    A DMA-based UART firmware that implements the NMEA 0183 GPS protocol, incrementally parsing sentences (e.g., GGA, RMC) into a structured GPS state.
    \item 
    \emph{Stepper Controller.}
    A standalone three-axis stepper controller that exposes a compact ASCII command protocol over UART and computes acceleration/braking profiles to configure timer-driven step pulses.
    \item 
    \emph{CNC (Grbl-based controller).}
    A Cortex-M port of the Grbl motion-control stack: it receives G-code over serial, plans linear and circular moves, and drives three stepper axes via timer-based step/direction pulses.
    \item 
    \emph{Soldering Station.}
A TS100-derived soldering iron controller with PID-based tip temperature control, sleep/boost modes, and PWM-driven heating.
\end{itemize}

All four targets are bare-metal firmware from open-source projects, ported to the STM32L476RG with minimal changes (e.g., pin remapping, disabling unused peripherals) while preserving core parsing and control logic. All run on the same MCU and are fuzzed over a serial interface, but span diverse behaviors: (i) command/protocol parsing (G-code, NMEA, custom ASCII), (ii) continuous control loops (PID and stepper profiles), and (iii) DMA/interrupt-driven I/O with timer-based actuation. EM-guided selection improves coverage across all targets compared to the random baseline, suggesting the approach generalizes across common MCU patterns rather than being tailored to a single program. While we evaluate only bare-metal control firmware, the method is application-agnostic and should extend to RTOS and networked firmware, which we leave to future work. Because we rebuild all targets for STM32L4 with our toolchain, instruction/basic-block counts may differ from prior reports; coverage is computed on these rebuilt binaries.

\begin{figure}[ht]
  \centering
  \includegraphics[width=0.5\columnwidth]{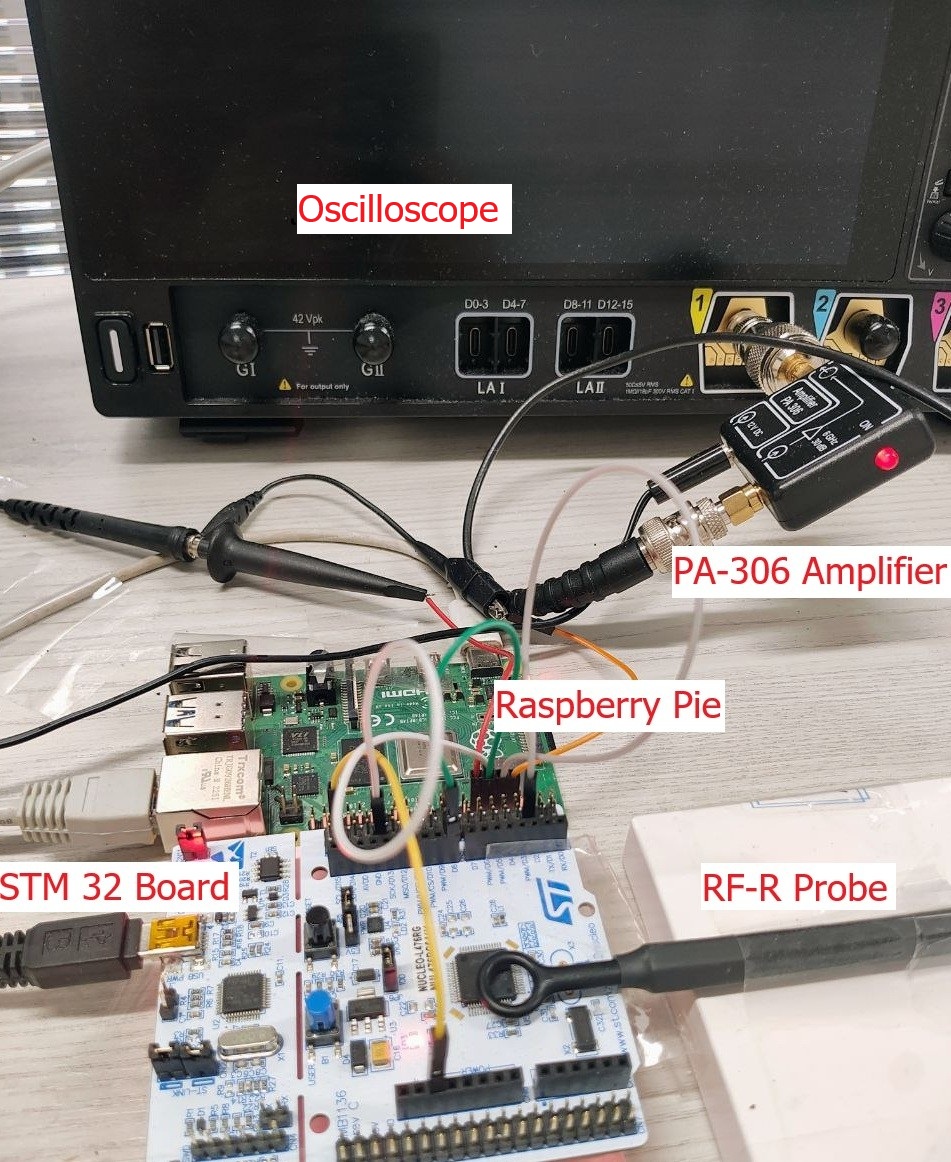}
    \caption{Hardware setup.}
  \label{fig:probe-setup}
\end{figure}

Fig.~\ref{fig:probe-setup} shows the experimental setup we used to collect the electromagnetic side channel traces.
The inputs are sent over UART from a Raspberry~Pi bridge, which simultaneously triggers EM capture on a Rigol MHO5104.
The signal is measured with an RF-R 50-1 H-field probe positioned at the MCU’s rear-left corner, where the measurment has the highest SNR. 
The probe output is amplified using a PA-306 with a 30 dB gain across 100 kHz–6 GHz and recorded on channel 1 of a Rigol MHO5104, which features a 50~$\Omega$ input, over a millisecond–scale window that spans reception and execution. 
Using deep memory (25 M points) avoids on-instrument decimation, and the resulting raw 8-bit waveforms are transferred to the host for offline processing.

\subsection{Evaluation Results}

\subsubsection{Guidance Effectiveness}

\begin{figure*}[ht]
  \centering
  \setlength{\tabcolsep}{0pt}
  \begin{tabular}{@{}c c c c c@{}}
\rotatebox[origin=c]{90}{\footnotesize\# of instruction hits}&    \begin{minipage}{0.25\textwidth}\centering
      (a) GPS\\[2pt]
      \includegraphics[width=\linewidth]{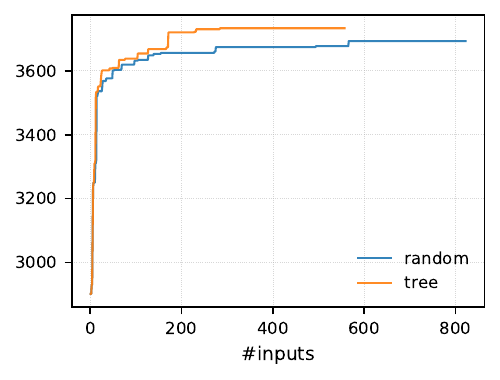}
    \end{minipage} &
    \begin{minipage}{0.25\textwidth}\centering
      (b) Stepper\\[2pt]
      \includegraphics[width=\linewidth]{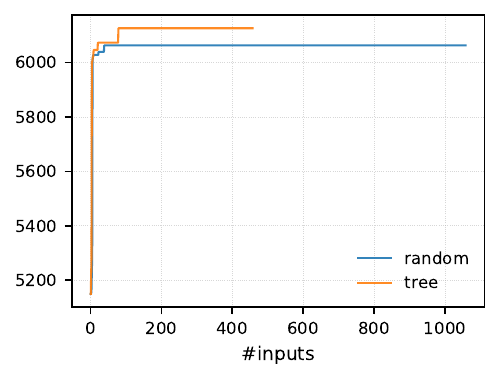}
    \end{minipage} &
    \begin{minipage}{0.25\textwidth}\centering
      (c) CNC \\[2pt]
      \includegraphics[width=\linewidth]{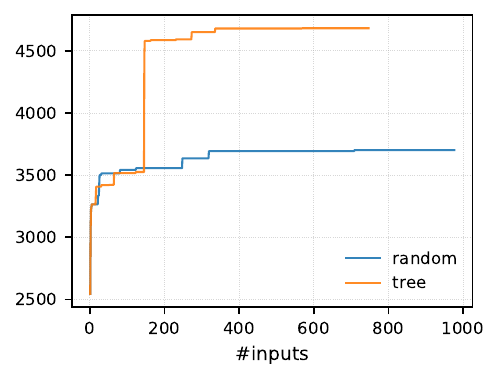}
    \end{minipage} &
    \begin{minipage}{0.25\textwidth}\centering
      (d) Soldering Station\\[2pt]
      \includegraphics[width=\linewidth]{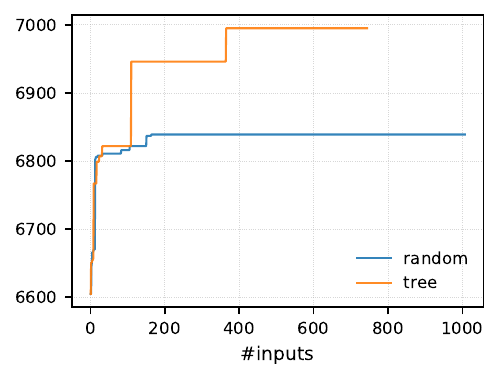}
    \end{minipage}
    \\[6pt]

    \rotatebox[origin=c]{90}{\footnotesize\# of edge hits} &
    \begin{minipage}{0.25\textwidth}\centering
      \includegraphics[width=\linewidth]{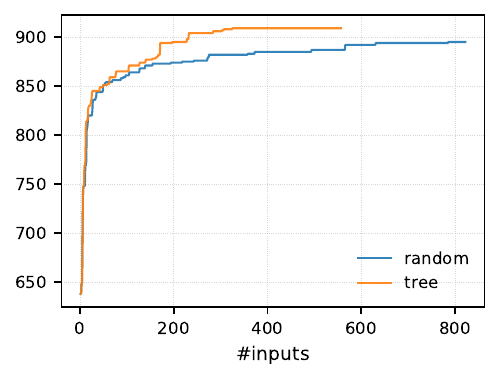}
    \end{minipage} &
    \begin{minipage}{0.25\textwidth}\centering
      \includegraphics[width=\linewidth]{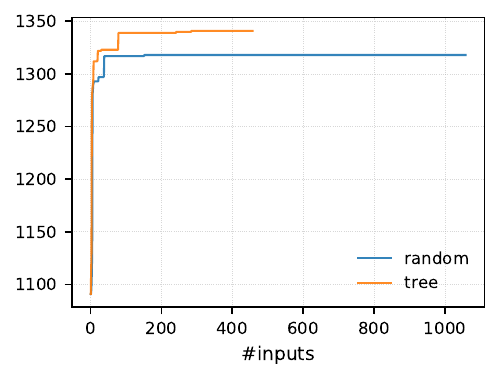}
    \end{minipage} &
    \begin{minipage}{0.25\textwidth}\centering
      \includegraphics[width=\linewidth]{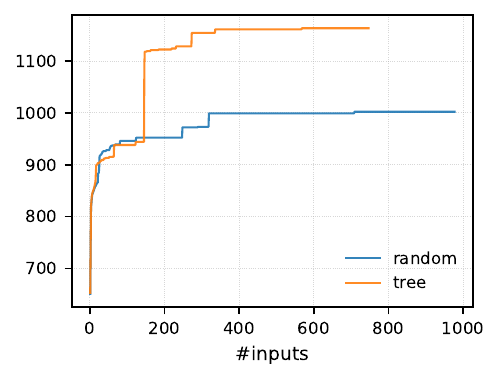}
    \end{minipage} &
    \begin{minipage}{0.25\textwidth}\centering
      \includegraphics[width=\linewidth]{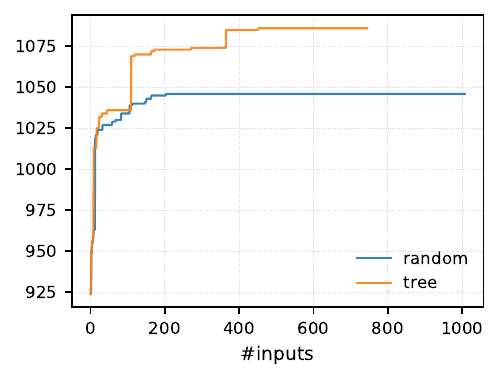}
    \end{minipage}
  \end{tabular}
  \caption{Coverage growth vs.\ inputs for EM-guided vs.\ Random selections. Y-axes differ across subplots.}
  \label{fig:cov-curves}
\end{figure*}

To compare EM-guided selection against random selection, we plot how code coverage grows with the number of executed inputs for both modes.
For each run, we record—input by input—the number of previously unseen instructions and edges first reached by that input.
Figure~\ref{fig:cov-curves} shows cumulative instruction and edge coverage as fuzzing progresses for each firmware.
Across the evaluated platforms, EM-guided selection exceeds the final instruction and edge coverage of random selection and typically reaches that coverage with fewer inputs under the same mutation operators, initial seeds, and input budget.
This indicates that EM-guided selection is more efficient at discovering reachable code than unguided selection.
In the remainder of this section, we quantify the coverage growth speed and final coverage using summary metrics derived from the same runs.

Table~\ref{tab:instr-coverage} summarizes the total number of distinct instructions covered by each method and the instructions covered exclusively by one method. 
These unique instructions indicate small regions of code that are only exercised by a single method.
Across all four firmware targets, EM-guided selection achieves higher total instruction coverage. In the CNC and Soldering Station, it also discovers a substantially larger set of unique instructions than random selection, showing that EM-guided scheduling not only accelerates coverage growth but also exposes code regions that unguided fuzzing never reaches. In Stepper and GPS, the random mode has a small number of unique instructions, but the net unique coverage still favors EM-guided selection.

\begin{table}[ht]
    \centering
    \caption{Total instruction coverage and unique instructions}
    \label{tab:instr-coverage}
      \resizebox{\columnwidth}{!}{
    \begin{tabular}{llrr}
        \toprule
        Firmware & Mode & Total instr & Unique instr \\
        \midrule
        Stepper & Random selection & 6063 & 17 \\
                &  EM-guided    & 6126 & 80 \\
        \midrule
        CNC     & Random selection & 3701 & 8 \\
                & EM-guided    & 4682 & 989 \\
        \midrule
        GPS     & Random selection & 3693 & 16 \\
                &  EM-guided    & 3733 & 56 \\
        \midrule
        Soldering Station  & Random selection & 6839 & 0 \\
                &  EM-guided    & 6995 & 156 \\
        \bottomrule
  \end{tabular}
  }
\end{table}

To quantify how quickly coverage grows, we measure how many inputs each method requires to reach the same instruction and edge coverage. Table~\ref{tab:baseline-speed} reports, for each target, the number of inputs needed to reach the final coverage attained by random selection. In every case, EM-guided selection reaches this baseline with substantially fewer inputs for both instruction and edge coverage, indicating faster discovery of reachable code.

\begin{table}[ht]
\centering
\caption{Baseline-coverage efficiency: number of inputs required for each mode to reach \textsc{Random}’s final cumulative instruction and edge coverage.}
  \label{tab:baseline-speed}
  \resizebox{\columnwidth}{!}{
\begin{tabular}{llrrr}
\toprule
Firmware & Mode &
  \multicolumn{1}{c}{Baseline instr} &
  \multicolumn{1}{c}{Baseline edges}  \\
\midrule
GPS     & Random selection& 566 & 786   \\
     & EM-guided   & 171 & 232   \\
\midrule
Stepper & Random selection &   39 & 152 \\
 &  EM-guided               &  21 & 21   \\
\midrule
CNC     & Random selection & 709 & 709  \\
     &  EM-guided   & 146 & 146   \\
\midrule
Soldering Station & Random selection & 164 & 204  \\
     &  EM-guided   & 110 & 110   \\
\bottomrule
\end{tabular}
}
\end{table}

For EM-guided selection, we distinguish between inputs that are inserted as new nodes in the divergence tree and inputs that the tree rejects as behaviorally redundant. Only the node-inserting inputs influence future scheduling decisions. For evaluation, however, we archive the discarded mutations and replay them separately in QEMU to check whether any of them exercise instructions or edges that are not already covered by the tree-selected inputs. In our experiments, the discarded inputs do not reveal additional coverage beyond what is reached by the divergence tree, indicating that the discarded nodes do not hide extra behavior.

\subsubsection{Novelty Efficiency}
To compare the novelty efficiency of different methods, we measure how often each mode, EM-guided selection and random selection, produces inputs that exercise previously unseen behaviors. For each generated input, we replay it in QEMU and mark it as novel if it reveals at least one new instruction or edge relative to all prior inputs from the same mode. We then compute the novelty hit-rate, defined as the fraction of generated inputs that are novel under each selection mode. From the perspective of our method, any input that appears behaviorally novel at the time domain level but does not reveal new coverage is a coverage-level false positive. A Higher novelty hit-rate, therefore, indicates not only better exploration but also fewer such false positives: more of the EM-guided novel inputs correspond to genuinely new instructions or edges, whereas a low hit-rate indicates that the scheduler spends much of its budget on redundant inputs that do not expand coverage. This metric isolates the effect of the scheduling policy. Mutation operators, number of mutations in the havoc stage, and initial seeds are identical across modes, so differences in novelty reflect only the choice of which seed is mutated next.

\begin{table}[ht]
  \centering
  \caption{Novelty Efficiency: \textsc{EM-guided selection} vs. \textsc{Random Selection}.}
\label{tab:novelty}
 \resizebox{\columnwidth}{!}{
\begin{tabular}{llccc}
\toprule
Firmware & Mode & Instr (\%) & BBLs (\%) & Edges (\%) \\
\midrule
GPS     & Random Selection & 3.41 & 3.41 & 4.74 \\
        & EM-guided   & 4.85 (1.42$\times$) & 4.85 (1.42$\times$) & 6.82 (1.44$\times$) \\
\midrule
Stepper & Random Selection & 0.85 & 0.85 & 1.32 \\
        & EM-guided        & 2.19 (2.58$\times$) & 2.19 (2.58$\times$) & 3.28 (2.48$\times$) \\
\midrule
CNC     & Random Selection & 1.89 & 1.89 & 4.38 \\
        & EM-guided   & 2.54 (1.34$\times$) & 2.54 (1.34$\times$) & 5.22 (1.19$\times$) \\
\midrule
Soldering Station & Random Selection & 1.29 & 1.29 & 2.09 \\
\bottomrule
\end{tabular}
}
\end{table}

Table ~\ref{tab:novelty} compares EM-guided selection with random selection. EM-guided selection yields a substantially higher novelty hit rate, showing that inputs chosen by our strategies are more likely to explore previously unseen executions than those chosen at random. The results confirm that EM-guided selection drives the firmware into new behaviors more frequently under the same mutation budget.
\section{Related Work}

Traditional coverage-guided fuzzers recover coverage feedback from the firmware binary~\cite{fioraldi2022libafl}. A common technique is rehosting, where the extracted binary runs in an emulator and coverage is collected by instrumenting translated basic blocks in the emulated CPU~\cite{zaddach2014avatar, zheng2019firm}. Embedded firmware depends on peripheral behavior, interrupts, and timing, so incomplete models or timing mismatches can block execution and limit the usefulness of the recovered feedback. 
Rehosting frameworks therefore add mechanisms to handle missing device behavior.

A common direction is to abstract peripherals rather than faithfully emulate concrete devices. 
Some systems bypass device-specific logic by intercepting hardware abstraction layer (HAL) interactions and replacing them with high-level stubs inferred from binary analysis or library matching~\cite{clements2020halucinator}. 
Others keep the firmware image intact and synthesize peripheral behavior at runtime using simulated peripherals, documentation-derived interface contracts, and DMA-aware abstractions~\cite{gustafson2019toward, feng2020p2im, mera2021dice}. 
While these strategies can keep execution progressing, they remain sensitive to undocumented register semantics, complex device interactions, and asynchronous events.

To overcome the limitations of pre-built peripheral modeling, rehosting systems employ symbolic execution to supply missing device behavior. 
For example, tools like µEmu~\cite{zhou2021automatic} and Fuzzware~\cite{scharnowski2022fuzzware} generate values for memory-mapped I/O and interrupt-driven state, preventing execution from stalling on unresolved MMIO reads and interrupt-dependent control flow. 
This approach scales poorly in MMIO-heavy firmware where symbolic values propagate through global state. 
Furthermore, some techniques approximate timing via scheduling heuristics and manual clock advancement~\cite{scharnowski2023hoedur}. 
While these can keep simple firmware running, they frequently break when behavior depends on precise timing relationships.

Different from rehosting, static and symbolic analyses reason about the firmware code directly using techniques such as dependence analysis, slicing, taint tracking, and symbolic reasoning. 
These methods can expose deep logic flaws, but they assume white-box access and do not provide a lightweight, per-execution feedback signal suitable for guiding large-scale mutation-based exploration~\cite{shoshitaishvili2015firmalice, redini2020karonte}.

State-of-the-art blackbox fuzzers~\cite{chen2018iotfuzzer, lee2015fuzzing} operate with no knowledge of the target’s internals and rely on external monitors to flag failures, exceptions, or crashes.
However, these approaches either have a limited scope or require expensive manual reverse engineering, and consequently do not provide generic, automated tooling that can be transferred across targets.

Recent work has explored using side channels to approximate program coverage or detect behavioral changes. Sperl et al.~\cite{sperl2019side} use power traces with machine-learning classifiers to reconstruct basic blocks and branch distances and to infer coverage with high correlation.
However, this requires per-device calibration and a training phase that labels instruction types on the target, which breaks the black-box assumption and ties the method to a hardware-specific setup.

A different line of work~\cite{dunne2022powertrace} avoids control flow graph reconstruction and training by applying statistical distance tests to averaged power traces,  flagging inputs whose traces differ from a reference profile. 
This enables black-box detection of hidden modes in crafted firmware, but the requirement to average thousands of traces per input to filter noise and isolate the deterministic control-flow signal severely limits throughput. 
Additionally, the method exhibits a blind spot for short, compute-heavy paths with little memory or peripheral activity, biasing exploration toward memory-intensive behaviors. 

Diffuzz~\cite{nilizadeh2019diffuzz} maximizes observable execution differences but relies on instrumentation or explicit measurement hooks (e.g., memory-usage signals), again assuming control over the binary/runtime. Dong et al.~\cite{dong2025emfuzz} introduce EMFuzz, which mutates IEMI parameters and uses ML to classify actuator-level effects via sensors; this provides black-box feedback about actuator-level malfunction rather than program-path exploration, which is the focus of our work.

\section{Conclusion}
This work demonstrates that electromagnetic side-channel measurements provide a practical feedback signal for fuzzing firmware in fully black-box settings where binaries cannot be extracted, instrumented, or reliably emulated. Using EM-derived divergence as guidance, our fuzzer explores new behaviors more efficiently: across multiple real targets, it achieves higher code coverage and reaches deeper paths with fewer inputs than unguided fuzzing. These results show that effective fuzzing guidance can be recovered directly from physical side-channel signals.

\section{Acknowledgment}
This work is based upon the work supported by the National Center for Transportation Cybersecurity and Resiliency (TraCR) (a U.S. Department of Transportation National University Transportation Center) Headquartered at Clemson University, Clemson, South Carolina, USA. 
Any opinions, findings, conclusions, and recommendations expressed in this material are those of the author(s) and do not necessarily reflect the views of TraCR, and the U.S. Government assumes no liability for the contents or use thereof.
In addition, Ziming Zhao is supported in part by the National Science Foundation under grant numbers 2508320 and 2512972.

\bibliographystyle{IEEEtran}
\bibliography{Section/reference}

\appendices

\end{document}